\documentclass{article}

\usepackage{neurips_2021}
\usepackage[dvipdfmx]{graphicx}
\usepackage[utf8]{inputenc} % allow utf-8 input
\usepackage[T1]{fontenc}    % use 8-bit T1 fonts
\usepackage{hyperref}       % hyperlinks
\usepackage{url}            % simple URL typesetting
\usepackage{booktabs}       % professional-quality tables
\usepackage{amsfonts}       % blackboard math symbols
\usepackage{nicefrac}       % compact symbols for 1/2, etc.
\usepackage{microtype}      % microtypography
\usepackage{xcolor}         % colors

\title{Artificial Perception Meets Psychophysics, Revealing a Fundamental Law of Illusory Motion}

\author{%
Taisuke Kobayashi\\
Laboratory of Neurophysiology\\
National Institute for Basic Biology\\
Okazaki, Japan\\
\texttt{taisukekbys@gmail.com} \\
\And
Eiji Watanabe\\
Laboratory of Neurophysiology\\
National Institute for Basic Biology\\
Okazaki, Japan\\
Department of Basic Biology\\
The Graduate University for Advanced Studies (SOKENDAI)\\
Miura, Japan \\
\texttt{eiji@nibb.ac.jp, eijwat@gmail.com} \\
}

\begin{document}

\maketitle

\begin{abstract}
``Rotating Snakes'' is a visual illusion in which a stationary design is perceived to move dramatically. In the current study, the mechanism that generates perception of motion was analyzed using a combination of psychophysics experiments and deep neural network models that mimic human vision. We prepared three- and four-color illusion-like designs with a wide range of luminance and measured their strength of induced rotational motion. As a result, we discovered the fundamental law that the effect of the four-color snake rotation illusion was successfully enhanced by the combination of two ``perceptual motion vectors'' produced by the two ``three-color'' designs. In years to come, deep neural network technology will be one of the most effective tools not only for engineering applications but also for human perception research.
\end{abstract}

\section{Introduction}
In peripheral vision, visual motion illusions have been widely known as the Fraser–Wilcox illusion (Fraser–Wilcox illusion, FWI) [6], or its variants [5, 15, 24] (Figure 1). These illusory designs consist of circular repetitions of stationary stimuli composed of segments with differing luminance that induces perception of rotational motion. The direction of illusory motion depends on the order of the luminance gradients. Among these variants, the Rotating Snakes illusion (RSI), design art by Prof. Akiyoshi Kitaoka, is particularly well-known for its dramatic illusion of visual motion [1, 13, 15, 18, 19]. Compared to the original design of the FWI, the RSI is always perceived in the expected direction, and the strength of perceived illusory motion is more significant. The RSI design is composed of four adjacent regions of different luminance; black, dark gray, white, and light gray, and the perceived illusory motion occurs in that order. The colored design, where dark gray is represented in blue and light gray in yellow, is the most noted and impressive RSI [14].

\begin{figure}[ht]
\centering
\includegraphics[keepaspectratio]{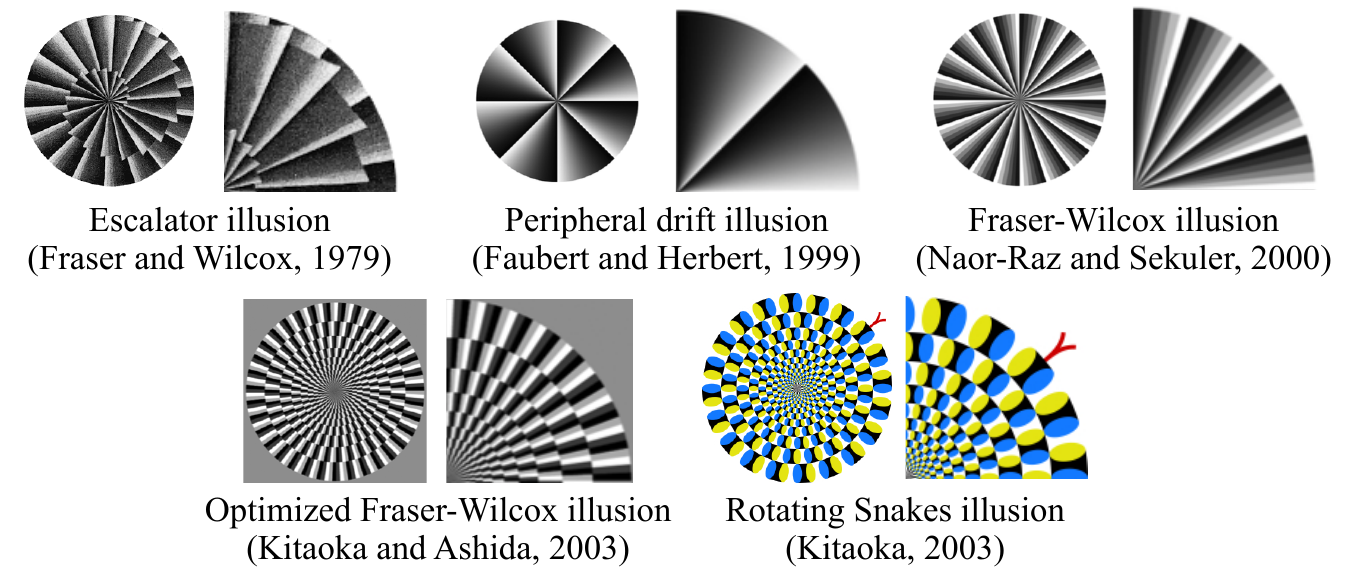}
\caption{Visual motion illusions in peripheral vision. Since the escalator illusion reported by Fraser and Wilcox (the original Fraser–Wilcox illusion), various motion illusions have been created. These illusory designs induce perception of rotational motion by circularly repeating stimuli composed of segments with luminance gradients.}
\end{figure}

One explanation for perception of illusory motion is that differences in adjacent contrasts produce differences in neural processing rates, in turn producing neural responses like those of physically moving stimuli [2, 3, 5, 24]. In another explanation for perception of illusory motion, Murakami et al. [23] proposed that the visual field’s motion generated by microsaccade, when combined with asymmetric retinal images produced by illusory designs, is converted into motion signals. Although these two hypotheses can coexist and seem to explain the mechanism of the illusion well, they do not seem to provide clear explanation for the strength of illusory movements unique to the RSI.

The two illusions, FWI and RSI, have noticeable differences in representation of luminance gradient (Figure 2). When luminance is aligned along the illusion’s major motion direction, luminance of the FWI rises in linear proportion from black to white, and then falls vertically from white to black. In contrast, the RSI’s gradient of luminance has a complex shape in which luminance rises curvilinearly to white across blue (dark gray), which is close to black luminance, and then falls curvilinearly from white to black across yellow (light gray), which is close to white luminance. The secret of the RSI seems to be hidden in its special shape of luminance gradient. Therefore, we attempted to investigate comprehensively the illusory motion of designs with various luminance gradients using colors in a wide range of hues.

Studies in neuroscience [26], visual perception [8, 17], and visual illusion [e.g., 9, 10, 21] using deep neural network (DNN) models are becoming more active and beginning to yield results. Therefore, we first explored a rough hypothesis with a deep learning model that mimics human vision [28] and then verified the hypothesis with psychophysics experiments.

\begin{figure}[b]
\centering
\includegraphics[keepaspectratio]{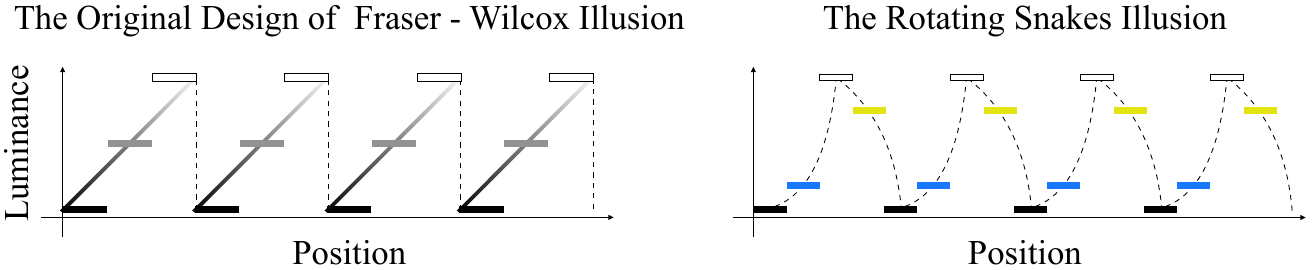}
\caption{Representation of luminance gradient of the two motion illusions. The vertical axis is luminance, and the horizontal axis is the spatial position of luminance. The major illusory motion occurs from left to right on the horizontal axis. The shape of the luminance gradient is curvilinear in the Rotating Snakes illusion but linear in the original Fraser–Wilcox illusion.}
\end{figure}

\section{Results}
\subsection{Motion illusion-like discrete designs (MIDDs)}
Motion illusion-like discrete designs (MIDDs) consisting of 3-color elements represented by black, intermediate luminance color (3C), and white and those consisting of 4-color elements represented by black, intermediate luminance color 1 (4C1), white, and intermediate luminance color 2 (4C2) were prepared (Figure 3). We then investigated the strength of illusory rotational motion generated in MIDDs by varying 3C, 4C1, and 4C2 in DNN model experiments and human psychophysics experiments.

\begin{figure}[ht]
\centering
\includegraphics[keepaspectratio]{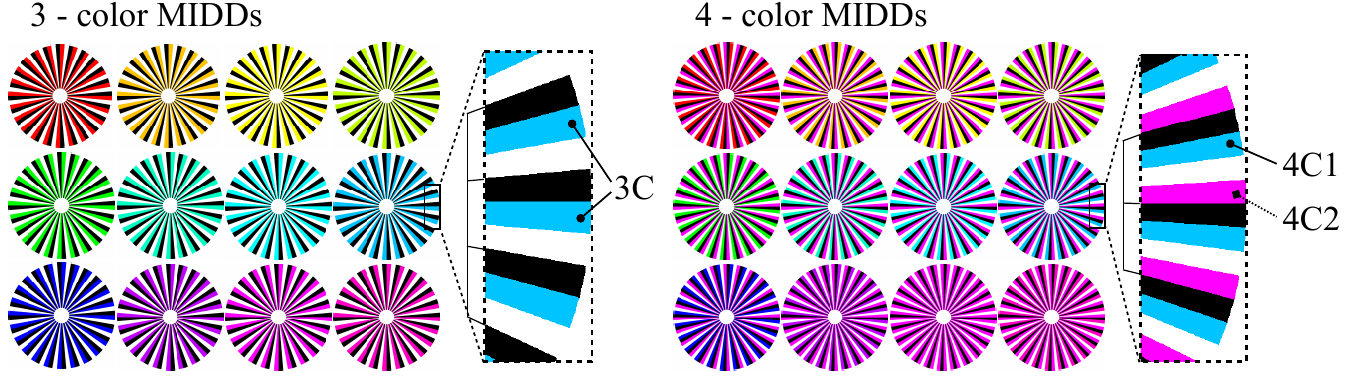}
\caption{Motion illusion-like discrete designs (MIDDs). Shown here are 24 examples (12 3-color and 12 4-color MIDDs) of visual stimuli used in DNN model experiments and psychophysical experiments. Three-color MIDDs consist of 3 elements represented by black, intermediate luminance color 1 (3C), and white; 4-color MIDDs consist of 4-color elements represented by black, intermediate luminance color 1 (4C1), white, and intermediate luminance color 2 (4C2).}
\end{figure}

\subsection{DNN model experiments}
We first calculated predicted illusory flows for each MIDD using a DNN model generated as a human vision simulator [28]. The network structure of this study’s model implemented [20] prediction hypotheses derived from the physiology and anatomy of the cerebral cortex and psychophysics [7, 12, 25, 27]. It was trained with first-person view videos, and as a result, it successfully simulated human illusory perception induced by the RSI [28]. Subsequent studies have reported that it reproduced a wide variety of perceptual phenomena concerning motion illusions other than the RSI [16]. Moreover, it has been reported to simulate not only the motion illusion but also the flash-lag effect and a kind of subjective contour [21].

DNN model experiments were conducted using 3-color MIDDs. The same 20 images copied from one of the MIDDs were input in sequence to the DNN model, two time-sequential predicted images were output, and the optical flow between the two images was calculated. Figures 4a and 4b show average values of optical flows in the rotational direction obtained from the DNN model’s predicted images. In the graph with hue as the horizontal axis, average rotational speed repeatedly rose and fell, but there was no clear correlation between hue and speed. Converting hue to luminance and plotting luminance as the abscissa, hues 60 and 240, which had the lowest and highest luminance values, were predicted to have motions with large rotational velocities in different directions, and the predicted motions’ magnitude appeared to be correlated with luminance. However, if we focused on the middle range of the two luminances above, the correlation was disturbed ($R^2$ of linear approximation $= 0.1428$, $y = -0.3376x + 0.3008$, dotted line in the upper right graph). Next, experiments using 4-color MIDDs were performed (Figures 4c and 4d). All graphs plotted with hue on the horizontal axis had similar shapes to the 3-color MIDDs and showed large ups and downs along the hue. As in the 3-color MIDDs, there appeared to be a linear correlation between the luminance of 4C2 and the rotation velocity for all 4C1. However, its correlation was very weak (mean $R^2$ of linear approximation $= 0.2099$, Std $= 0.0901$).

\begin{figure}[ht]
\centering
\includegraphics[keepaspectratio]{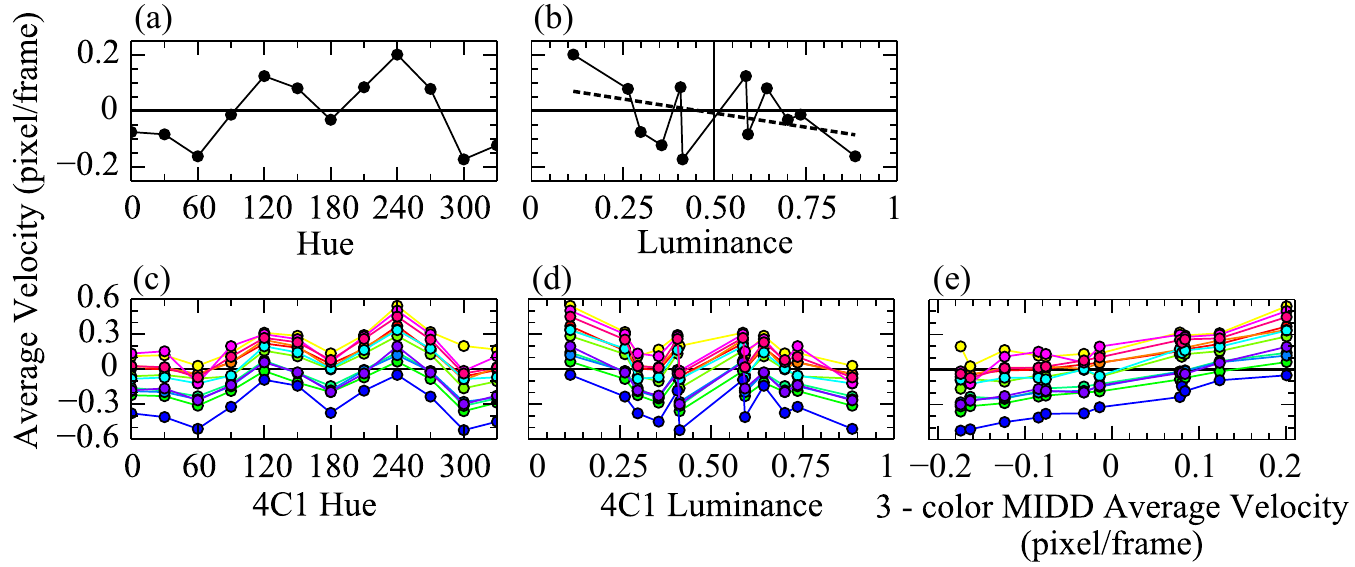}
\caption{DNN model experiments using 3-color (a and b) and 4-color MIDDs (c and d). In (a) or (c), the horizontal axis is a hue of 3C (a) or 4C1 (c), and the vertical axis is the average of detected rotational velocity. In (b) or (d), the horizontal axis is luminance of 3C (b) or 4C1 (d). In 4-color MIDDs, graphs were plotted in different colors for each hue of 4C2. See Figure 5 for the correspondence between plotted color and hue. The dotted line in (b) is a linear approximation. Figure 4e shows the relationship of the results of both 3- and 4-color MIDDs. The rotational velocity of 3-color MIDDs was plotted on the horizontal axis, and the rotational speed of 4-color MIDDs on the vertical axis.}
\end{figure}

\begin{figure}[ht]
\centering
\includegraphics[keepaspectratio]{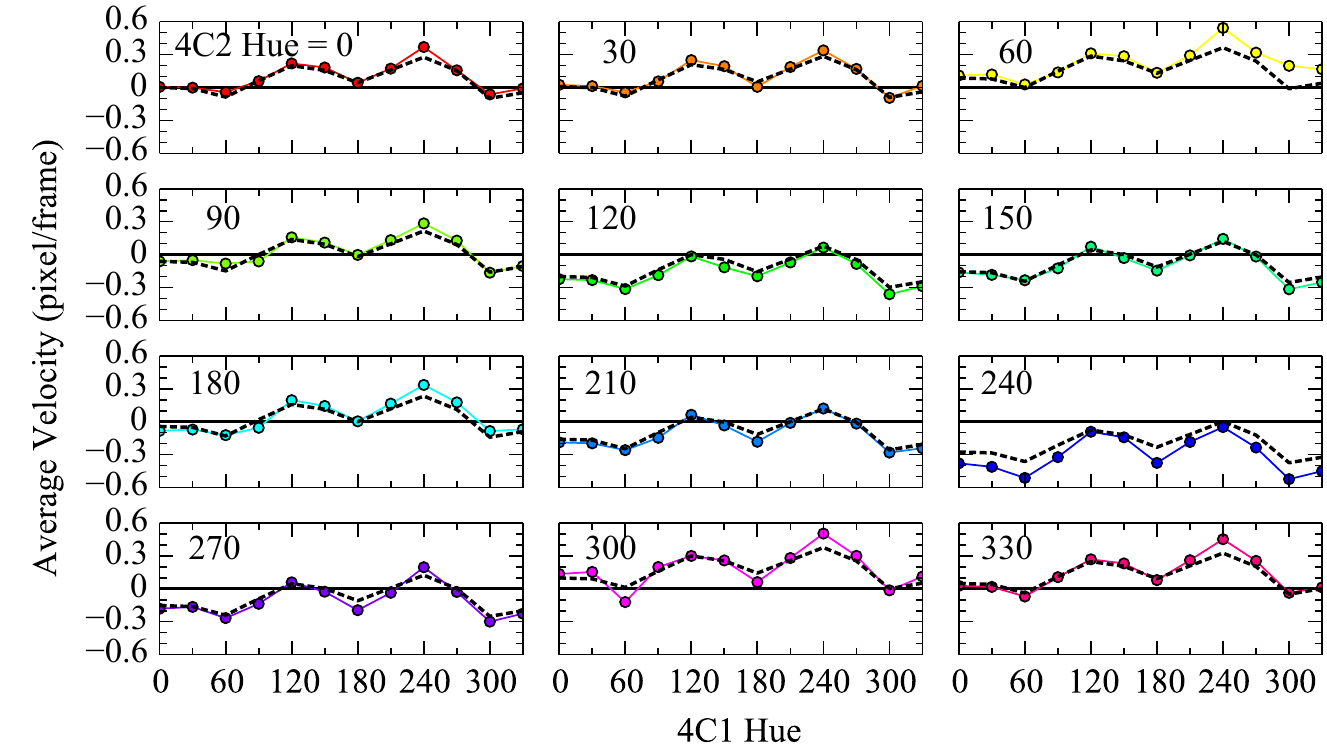}
\caption{Reconstruction of data of 4-color MIDDs from data of 3-color MIDDs (DNN model experiments). Data of 4-color MIDDs are shown by solid lines (same data as shown in Figure 4c), and simulated data from 3-color MIDDs are shown by dotted lines.}
\end{figure}

These experimental results suggested some linkage between data of 3- and 4-color MIDDs. Therefore, we plotted the rotational velocity of 3-color MIDDs on the horizontal axis and the rotational velocity of 4-color MIDDs on the vertical axis (Figure 4e); as expected, we observed positive correlation for all 4C1 (mean $R^2$ of linear approximation $= 0.9338$, std $= 0.0649$). Furthermore, the direction of rotation was switched near the point when 4C1 and 4C2 were in the same hue. This clearly suggested that the 4-color MIDD’s rotation velocity was determined by the 3-color MIDD’s characteristics. So next, if the 4-color MIDD was a simple composite of the 3-color MIDD, we simulated the 4-color MIDD’s data with those of the 3-color MIDD (Figure 5). As a result, strong correlation was found between simulated data and the original 4-color MIDD’s data (mean $R^2$ of linear approximation $= 0.8115$, Std $= 0.14685$).

\subsection{Psychophysics experiments}
Based on DNN experimental results, we designed psychophysics experiments using 3- and 4-color MIDDs as visual stimuli. To reduce the burden and the learning effect on subjects, the 4-color MIDDs’ stimuli were limited to a total of 24 with the hue of 4C2 fixed at 300.

\begin{figure}[ht]
\centering
\includegraphics[keepaspectratio]{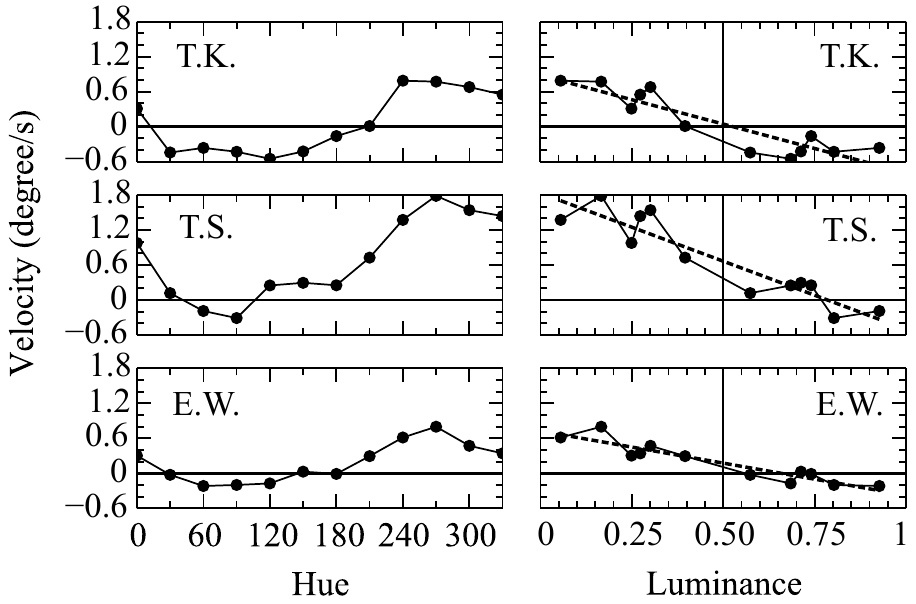}
\caption{Psychophysical experiments using 3-color MIDDs. Experimental results are shown for each of the three subjects. The figure’s notation is the same as in graphs of Figures 4a and 4b.}
\end{figure}

\begin{figure}[ht]
\centering
\includegraphics[keepaspectratio]{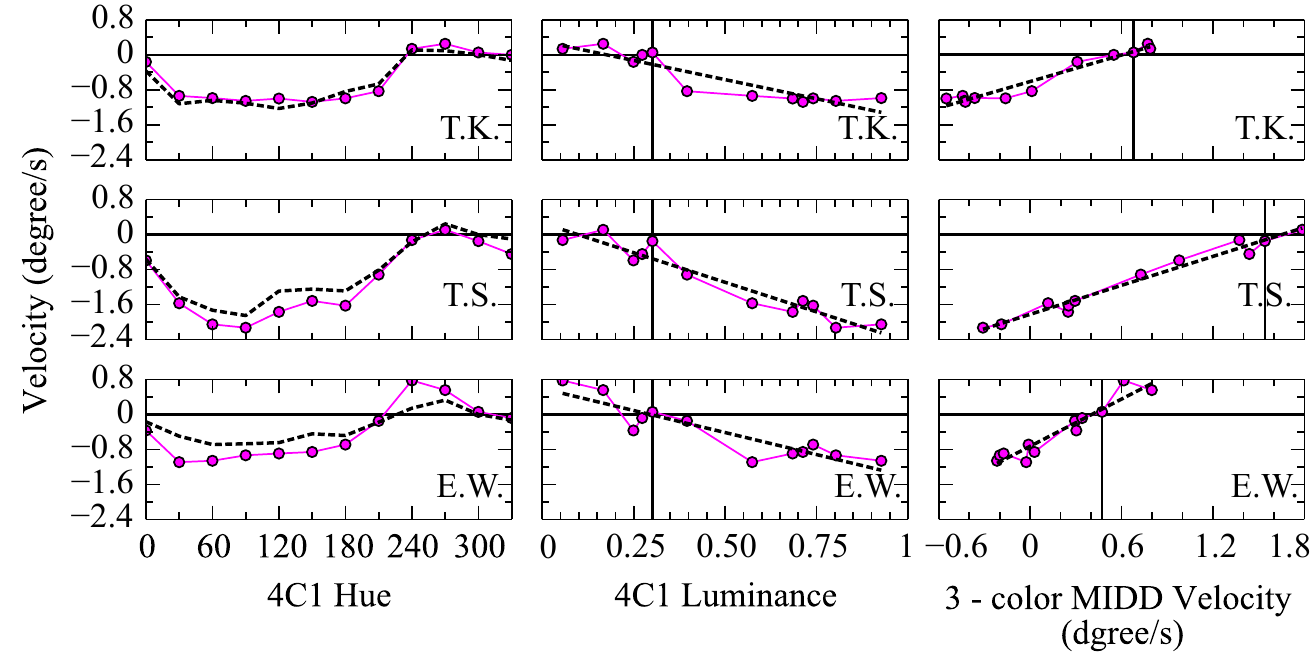}
\caption{Psychophysical experiments using 4-color MIDDs. Experimental results are shown for each of the three subjects. All vertical axes are rotational speeds obtained in 4-color MIDDs experiments. The horizontal axis of the three graphs on the left is hue, and the horizontal axis of the three graphs in the middle is luminance. The three graphs on the right show the relationship of both results of the 3- and 4-color MIDD experiments. Dotted lines in the left graphs mean rotational velocities simulated from data of 3-color MIDD experiments. Dotted straight lines in the middle and right graphs indicate linear approximations.}
\end{figure}

The 3-color MIDDs’ perceived rotational velocity is shown in Figure 6. With the hue plotted on the horizontal axis, the repetitive rise and fall of average rotational velocity was not observed as in the DNN model experiments, but rather an increase around hue 240 to 300. When luminance was plotted on the horizontal axis, clear correlation with rotational velocity was observed (T.K., $R^2$ of linear approximation $= 0.8156$, $y = -1.6735x + 0.8816$; T.S., $R^2 = 0.8533$, $y = -2.3341x + 1.8321$; E.W., $R^2 = 0.8707$, $y = -1.1013x + 0.7252$; dotted lines in right graphs), although there were individual differences. The direction reversed around luminance 0.5. In the DNN model experiment, luminance around 0.5 was greatly disturbed, differing from this psychological experiment’s data. However, the tendency at maximum and minimum luminances was the same in both experiments.

The 4-color MIDD’s rotational velocity is shown in Figure 7 (the left and middle graphs), and, as with the 3-color MIDD, there appeared to be strong correlation with luminance and detected rotational velocity (T.K., $R^2$ of linear approximation $= 0.8288$, $y = -1.7498x + 0.3055$; T.S., $R^2= 0.9180$, $y = -2.7080x + 0.2631$; E.W., $R^2 = 0.8188, y = -2.0145x + 0.5929$; dotted lines in the middle three graphs). As shown in the right three graphs in Figure 7, when the rotational velocity of the 4-color MIDD was plotted on the vertical axis and that of the 3-color MIDD on the horizontal axis, strong correlation was found (T.K., $R^2$ of linear approximation $= 0.9458$, $y = -1.0087x - 0.6141$; T.S., $R^2 = 0.9738$, $y = 1.1038x - 1.8237$; E.W., $R^2 = 0.9071$, $y = 1.7965x - 0.7277$; dotted lines in right graphs), and the switching point of the rotation direction was found near the luminance of hue 300 used in 4C2. When we plotted the 4-color MIDD’s rotational velocity, calculated using that of the 3-color MIDD, as in the DNN model experiment (dotted lines in the left three graphs in Figure 7), it was strongly suggested that the 4-color MIDD’s rotation was determined by the 3-color element’s rotation velocity (T.K., $R^2 = 0.9451$; T.S., $R^2 = 0.8752$; E.W., $R^2 = 0.3129$). Although rotational velocity differed by more than a factor of two between subjects, this trend did not change significantly in any subjects.

\section{Discussion}
\subsection{Personal differences}
A large difference in the strength of illusion was observed among subjects. Previous reports have indicated that individual differences in absolute strength might be due to individual differences in the eye movement’s magnitude (microsaccades) [23]. For all subjects, however, 3- and 4-color MIDDs showed strong correlation between strength of illusion and luminance; individual differences did not seem to change this tendency. In other words, the law found in this study seems to be robust, with few individual differences.

\subsection{Relationship between the shape of the luminance gradient and motion illusion}
In a single unit of 3-color MIDDs, the strength of illusion was lowest when the luminance of 3C is 0.5, and the strength of illusion tends to increase as it moves away from 0.5 (Figure 6, right). In other words, results suggested that the greater the curvature of the curve, the greater the strength of the illusion. This supports the result that the illusion’s strength was stronger when the gradient of luminance curved up and down than when it was linear up and down (Figure 2). How the curvilinear luminance gradient in space is converted into the strength of illusory motion is unclear. If we may venture a guess, reports have indicated that certain attributes with a curvilinear gradient, such as the Gabor filter [4, 22], are associated with visual information processing in the brain’s visual cortex, and this could be related to the present findings. As mentioned in the introduction, hypotheses focusing on luminance and contrast have been proposed [2, 3, 5, 24], but details of their shapes have not necessarily been studied. Further psychophysical and neurophysiological investigations focusing on the shape of luminance gradients are needed.

\subsection{Motion vector hypothesis}
The mechanism by which the RSI causes a large illusory motion, as inferred from this study’s results, is as follows (Figure 8).

\begin{figure}[t]
\centering
\includegraphics[keepaspectratio]{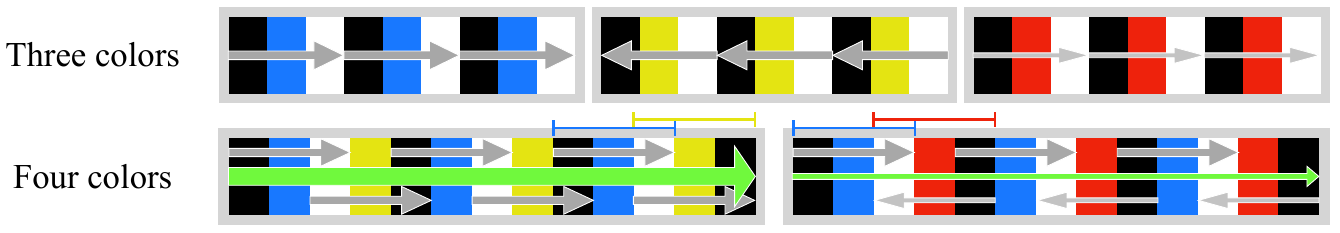}
\caption{Motion vector hypothesis. Arrows indicate the motion to be perceived. In the 3-colors condition, the unit generates a single motion. For example, when the intermediate luminance color is blue, motion is generated from black to white, and when the intermediate luminance color is yellow, motion is generated from white to black. For the 4-color condition, assume that vectors from the 3-color units overlap and that vectors are additive and subtractive. For example, if blue and yellow are used as intermediate luminance colors, the illusion’s assumed strength will be greater (a green arrow). If blue and red are used as intermediate colors, the direction of the two generated motions will be reversed, so the perceived motion will be smaller (a green arrow).}
\end{figure}

Four-color MIDDs can be regarded as representing a combination of two kinds of 3-color MIDDs. Assuming that one motion vector is generated by white, intermediate luminance color, and black, then one vector is generated by one unit in a 3-color MIDD, while two motion vectors are generated by using white and black of adjacent units in a 4-color MIDD. In other words, in 4-color MIDDs, white and black elements are involved in generating two kinds of motion vectors simultaneously, and the regions that generate motion vectors overlap spatially. In this sense, 4-color MIDDs might produce a larger illusion than 3-color MIDDs, or a smaller illusion, depending on the addition and subtraction of two motion vectors.

In the RSI, two motion vectors from the black-blue-white and white-yellow-black directions are expected to occur in the same direction. The two motion vectors overlap spatially and are not interrupted, so these motion vectors’ density is higher than that of the 3-color alone. In addition, both psychophysics and DNN model experiments have shown that blue (hue 240) and yellow (hue 60), which are close to the intermediate luminance colors used in the blue and yellow RSI (Figure 1), generate a large illusory motion, making yellow and blue one of the best choices for creating stimuli for 4-color MIDDs. A series of studies, starting with artificial perception experiments, seems to have backfilled ``the optimized Fraser–Wilcox illusions'' [13] created by the genius. On the other hand, the middle of the luminance gradient of the original design FWI units is close to luminance 0.5, where the illusory strength is presumed to be small. The direction of rotation of luminance 0.5 in Figure 6 (right) differs among the three subjects; this might be due to individual differences in the rotational direction of the original FWI paper.

\subsection{Improvement of DNN models}
Psychophysics experiments are ethically restricted because they require a great deal of time to obtain highly accurate data and because they place a heavy burden on subjects. Therefore, using simulator experiments with DNN models in psychophysics experiments is very effective. In addition, the simultaneous use of a simulator with psychophysics experiments can identify differences between DNN models and human perception and verify problems with cerebral and psychological theories on which DNN models are based. In fact, at some points in this study, the psychophysics experiments and the DNN model experiments showed different results.

The psychophysics experiments showed clear correlation between perceived rotation velocity and luminance (Figure 6, right; Figure 7, middle), but the DNN model experiments showed a large disturbance of the correlation around the middle luminance (Figures 4b and 4d). This might be a phenomenon related to the DNN model’s video input lacking explicit luminance information. In addition to cone cells that detect the three primary colors of red, green, and blue (RGB), the human retina has rod cells that detect light and dark, and it also has a much wider dynamic range of luminance than the RGB image’s input to DNNs. Peripheral vision with many rod cells has also been shown to play a decisive role in motion illusions’ generation [5, 6, 11, 24], and if this problem is solved, the DNN model could be upgraded to a more accurate visual simulator.

Artificial intelligence (AI) technology, represented by DNNs, is currently in a period of revolutionary development and is revolutionizing various fields, including that of psychology and understanding of human perception mechanisms. New technologies combined with psychological experiments create a great deal of room for innovative discoveries, and we can expect to contribute to the development of AI by providing feedback on these findings.

\section{Methods}
\subsection{Motion illusion-like discrete designs (MIDDs)}
The number of repetitions of units placed on a MIDD circle was set to 24, the same as stimuli used in the psychophysics experiments in [11] (Figure 3). Intermediate luminance colors were represented using the HLS color system, where L (lightness) = 0.5 and S (saturation) = 1 were fixed, and H (hue) was varied. Figure 3 shows examples of the stimuli. The hue varied from 0 to 330 degrees in 30-degree increments, resulting in 12 different hues used. The number of stimuli was 12 for the 3-color MIDDs and 144 for the 4-color MIDDs. They were processed to images 160 pixels wide and 120 pixels high for the DNN experiments. The circle’s center was set to the images’ center, and the circular design’s radius was set to 60 pixels. The circular design was processed to 191 pixels in radius for the psychophysics experiment.

\subsection{DNN model experiments}
DNN model experiments were performed in reference to the previous paper [28]. In brief, the connection weight model of trained DNN (PredNet) used in this study was identical to a 500K model described in the previous paper. PredNet is a DNN that learns to predict future image frames from time series data of images using mean square error. The model was obtained by 500K video frames training using first-person view videos (videos available to anyone as open source). The DNN model predicted the 21st image (P1 image) with reference to 22 consecutive images, which are 20 images copied from one test image. Next, the network predicted the 22nd image (P2 image) with reference to 21 consecutive images using the P1 image as the 21st image. Then, optical flow vectors between the P1 and P2 images were calculated with a customized Python program. Lucas-Kanade methods were utilized for optical flow analyses. In this study, we focused on the rotational motion occurring in the predicted image, and the center of the circular design drawn on the MIDD was assumed to be the center of rotation. Rotational velocity was calculated as the inner product of the optical flow vector and the unit vector indicating the direction of rotation around the circular design’s center calculated at the same position as the optical flow vector. To exclude noisy information unrelated to rotational motion, optical flow vectors, which are $|\cos{\theta}|<\frac{\sqrt{2}}{2}$ with respect to the unit rotation vector, were deleted. Optical flow vectors with a magnitude greater than the distance from the center of the circle were also removed.

\subsection{Calibration of display}
In the psychological experiment, we used a liquid crystal display (ASUS, PB287Q) to display stimuli. For the display to show the intended color information, gamma correction is necessary. Therefore, before the psychological experiment, we prepared to correct the RGB values to be displayed.

First, we measured the luminance when 8 increments of 256 shades of RGB were input (ColorCAL MKII, Cambridge Research Systems) (Supplemental Figure 1). In addition, the luminance of white (255,255,255) was measured to normalize luminance, and the luminance of black (0,0,0) was measured to obtain the offset.

Measured luminance does not have a linear relationship to RGB input values, and RGB values corresponding to the luminance for the RGB to be displayed must be input. The relationship between the RGB on the display (not the RGB used for input) and the luminance of each RGB at that time, $Y_{R}$, $Y_{G}$, and $Y_{B}$, is

\begin{center}
$Y_{R} = \frac{Y_{R_{Max}}}{255} R, Y_{G}=\frac{Y_{G_{Max}}}{255} G, Y_{B}=\frac{Y_{B_{Max}}}{255} B$
\end{center}

where $Y_{R_{Max}}$, $Y_{G_{Max}}$, and $Y_{B_{Max}}$ are luminances of R, G, and B at their maximum values, respectively. From this calculation, $Y_{R}$, $Y_{G}$, and $Y_{B}$ are obtained, and the RGB that matches the luminance is calculated from the measured values. The luminance of each hue of colors displayed using the RGB input values obtained by this procedure is shown in Supplemental Figure 2.

\subsection{Psychophysics experiment}
The psychophysics experiment conducted here was designed based on the method of Hisakata et al. [11] and with a program in Python using OpenGL. In this experiment, subjects were asked to answer whether they saw the stimuli rotating clockwise (CW) or counter-clockwise (CCW) by keyboard input, using a two-alternative forced choice. The subject’s face was fixed at 50 cm from the screen, and only the right eye was used for viewing. A gazing point with a viewing angle of 1 degree was set in the center of a white background, and a stimulus with an outer diameter of 7 degrees and an inner diameter of 1 degree was presented 12 degrees to the left of center for 0.5 seconds. The subject looked at the gazing point and viewed the stimulus with peripheral vision. When the subject responded, the next stimulus was played, but the stimulus was designed with a minimum of one second between presentation of the previous stimulus and playback of the next.

To examine the stimuli’s illusory motion quantitatively, stimuli were intentionally rotated, and the condition in which the illusory motion does not occur was calculated. Two types of stimuli were prepared: the original image and its left–right reversed version. This counteracts perceived rotational velocity bias. To obtain statistics on responses, we presented the same condition multiple times with randomly varying stimulus types and rotational velocities. From statistical data obtained through this procedure, we calculated the rotational velocity of the stimulus when the probability of answering that the stimulus was rotating in the CW and CCW directions was the same.

Supplemental Figure 3 shows graphs of the psychophysics experiment’s raw data. The horizontal axis is the velocity at which the stimulus was intentionally rotated, with CCW as the positive direction, and the vertical axis is the probability of responding that the stimulus was rotated to CCW for each stimulus. The obtained psychometric curves were fitted with a cumulative Gaussian function to calculate the rotational velocity and the rotational cancellation velocity when the probability became 0.5. Rotational cancellation velocity is the velocity required to cancel the presented image’s rotation and the direction of rotation due to the image’s motion illusion is the velocity multiplied by minus. Then, the rotation velocity of the stimulus as (static rotation velocity of the reversed stimulus) $-$ (static rotation velocity of the original stimulus) / 2 was calculated.

Subjects were the authors, T.K. and E.W., plus one naïve subject, for a total of three (all healthy subjects with normal vision). The experiment was conducted according to rules set by the ethics committee of the institution to which the author belongs. In 3- and 4-color MIDDs, intentional stimulus rotational velocities were set to $-$2.4 to $+$2.4$^{\circ}$/s and $-$5.0 to $+$5.0$^{\circ}$/s, respectively, and velocity intervals to 0.4$^{\circ}$/s and 1.0$^{\circ}$/s. The number of presentations of stimuli in the same condition was 24 and 20, respectively, and the number of presentations of all conditions for one stimulus was 624 and 440.

\subsection{Open source}
All program codes (DNN, optical flow analysis, and psychophysics stimulus presentation software), trained models, and stimulus images will be released as open source after the paper’s acceptance.

\newpage

\section*{References}
{
\small

[1]	Ashida, H., Kuriki, I., Murakami, I., Hisakata, R., \& Kitaoka, A. (2012). Directionspecific fMRI adaptation reveals the visual cortical network underlying the "Rotating Snakes" illusion. {\it NeuroImage} {\bf 61}, 1143--1152.

[2]	Backus, B. T., \& Oruç, I. (2005). Illusory motion from change over time in the response to contrast and luminance. {\it Journal of Vision} {\bf 5}, 1055--0169.

[3]	Conway, B. R., Kitaoka, A., Yazdanbakhsh, A., Pack, C. C., \& Livingstone, M. S. (2005). Neural basis for a powerful static motion illusion. {\it The Journal of Neuroscience} {\bf 25}, 5651--5656.

[4]	Daugman, J. G. (1980). Two-dimensional spectral analysis of cortical receptive field profiles. {\it Vision Research} {\bf 20}, 847--856.

[5]	Faubert, J., \& Herbert, A. M. (1999). The peripheral drift illusion: A motion illusion in the visual periphery. {\it Perception} {\bf 28}, 617--621.

[6]	Fraser, A., \& Wilcox, K. J. (1979). Perception of illusory movement. {\it Nature} {\bf 281}, 565--566.

[7]	Friston, K. (2005). A theory of cortical responses. {\it Philosophical Transactions of the Royal Society B Biological Science} {\bf 360}, 815--836.

[8]	Funke, C. M., Borowski, J., Stosio, K., Brendel, W., Wallis, T. S. A., \& Bethge, M. (2021). Five points to check when comparing visual perception in humans and machines. {\it Journal of Vision} {\bf 21}, 16.

[9]	Gomez-Villa, A., Martin, A., Vazquez-Corral, J., \& Bertalmio, M. (2019). Convolutional neural networks can be deceived by visual illusions. {\it Proceedings of the IEEE conference on computer vision and pattern recognition}, pp. 12309--12317.

[10]	Gomez-Villa, A., Martín, A., Vazquez-Corral, J., Bertalmío, M. \& Malo, J. (2020). Color illusions also deceive CNNs for low-level vision tasks: Analysis and implications. {\it Vision Research} {\bf 176}, 156--174.

[11]	Hisakata, R. \& Murakami, I. (2008). The effects of eccentricity and retinal illuminance on the illusory motion seen in a stationary luminance gradient. {\it Vision Research} {\bf 48}, 1940--1948.

[12]	Kawato, M., Hayakawa, H., \& Inui T. (1993). A forward-inverse optics model of reciprocal connections between visual cortical areas. {\it Network: Computation in Neural systems} {\bf 4}, 415--422.

[13]	Kitaoka, A. (2017). The Fraser-Wilcox illusion and its extension. A. G. Shapiro and D. Todorović (Eds.), {\it The Oxford Compendium of Visual Illusions, Oxford University Press}, pp. 500--511.

[14]	Kitaoka, A., Akiyoshi's illusion pages. Available at http://www.ritsumei.ac.jp/~akitaoka/index-e.html.

[15]	Kitaoka, A., \& Ashida, H. (2003). Phenomenal characteristics of the peripheral drift illusion. {\it Vision} {\bf 15}, 261--262.

[16]	Kobayashi, T., Kitaoka, A., Kosaka, M., Tanaka, K., \& Watanabe, E. (2022). Motion illusion-like patterns extracted from photo and art images using predictive deep neural networks. {\it Scientific Reports} {\bf 12}, 3893.

[17]	Kriegeskorte, N., (2015). Deep Neural Networks: A New Framework for Modeling Biological Vision and Brain Information Processing. {\it Annual Review of Vision Science} {\bf 1}, 417--446.

[18]	Kuriki, I., Ashida, H., Murakami, I., \& Kitaoka, A. (2008). Functional brain imaging of the Rotating Snakes illusion by fMRI. {\it Journal of Vision} {\bf 8}, 1--10.

[19]	Kubota, Y., Hayakawa, T., \& Ishikawa, M. (2021). Dynamic perceptive compensation for the rotating snakes illusion with eye tracking. {\it PLoS ONE} {\bf 16}, doi: 10.1371/journal.pone.0247937.

[20]	Lotter, W., Kreiman, G. \& Cox, D. (2016). Deep predictive coding networks for video prediction and unsupervised learning. arXiv:1605.08104.

[21]	Lotter, W., Kreiman, G. \& Cox, D. (2020). A neural network trained for prediction mimics diverse features of biological neurons and perception. {\it Nature Machine Intelligence} {\bf 2}, 210--219.

[22]	Marcelja, S. (1980). Mathematical description of the responses of simple cortical cells. {\it Journal of the Optical Society of America} {\bf 70}, 1297--1300.

[23]	Murakami, I., Kitaoka, A., \& Ashida, H. (2006). A positive correlation between fixation instability and the strength of illusory motion in a static display. {\it Vision Research} {\bf 46}, 2421--2431.

[24]	Naor-Raz, G., \& Sekuler, R. (2000). Perceptual dimorphism in visual motion from stationary patterns. {\it Perception} {\bf 29}, 325--335.

[25]	Rao, R.P., \& Ballard, D.H. (1999). Predictive coding in the visual cortex: a functional interpretation of some extra-classical receptive-field effects. {\it Nature Neuroscience} {\bf 2}, 79--87.

[26]	Richards, B.A., Lillicrap, T.P., Beaudoin, P. et al. (2019). A deep learning framework for neuroscience. {\it Nature Neuroscience} {\bf 22}, 1761--1770.

[27]	Watanabe, E., Matsunaga, W., \& Kitaoka, A. (2010). Motion signals deflect relative positions of moving objects. {\it Vision Research} {\bf 50}, 2381--2390.

[28]	Watanabe, E., Kitaoka, A., Sakamoto, K., Yasugi, M. \& Tanaka, K. (2018). Illusory motion reproduced by deep neural networks trained for prediction. {\it Frontiers in Psychology} {\bf 9}, doi: 10.3389/fpsyg.2018.00345.

}

\newpage
\renewcommand{\figurename}{Supplemental Figure}
\setcounter{figure}{0}

\begin{figure}[h]
\centering
\includegraphics[keepaspectratio]{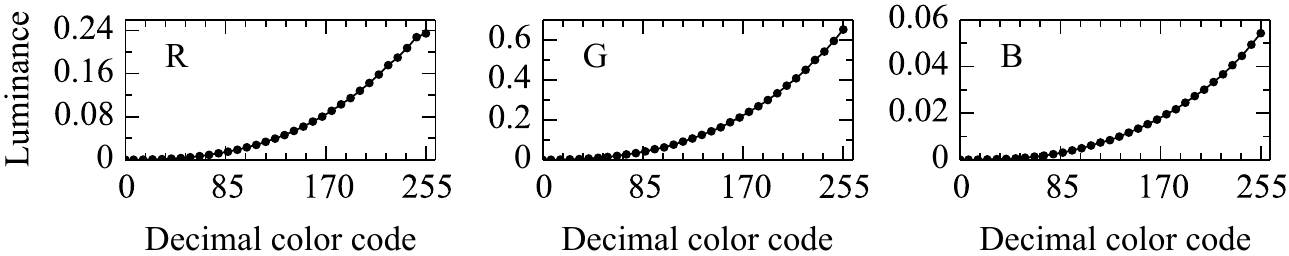}
\caption{The relationship between RGB and luminance values of the liquid crystal display before calibration. The horizontal axis indicates the decimal color codes (8-bit), and the vertical axis means luminance value. The luminance value measured by inputting the decimal color code (0,0,0) into the display was set to zero, and the luminance measured by inputting (255,255,255) was set to one, to normalize all luminance values.}
\end{figure}

\newpage

\begin{figure}[h]
\centering
\includegraphics[keepaspectratio]{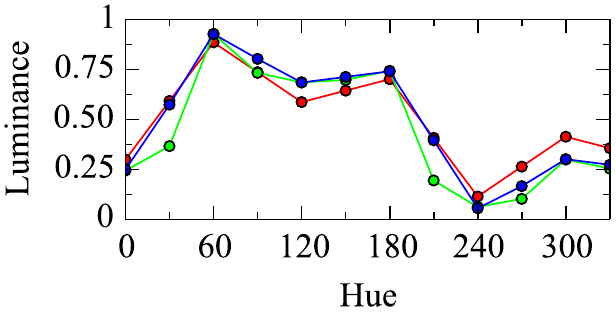}
\caption{Calibrating the display. The horizontal axis shows hue calculated from decimal color code, and the vertical axis indicates luminance value. Green: before calibration. Blue: after calibration. Red: theoretical values calculated from the decimal color codes.}
\end{figure}

\newpage

\begin{figure}[h]
\centering
\includegraphics[keepaspectratio]{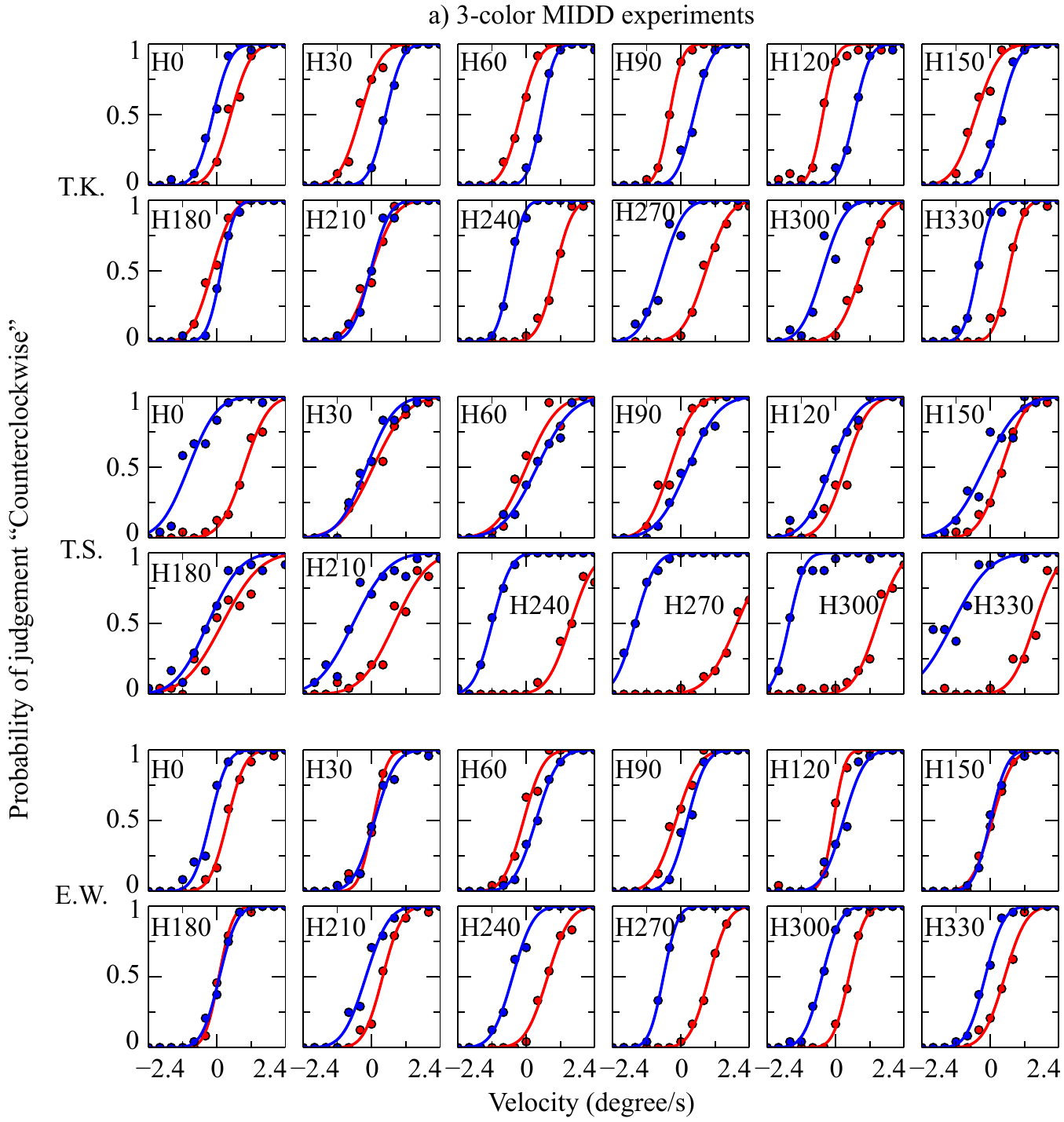}
\renewcommand{\thefigure}{3a}

\end{figure}

\newpage

\begin{figure}[h]
\centering
\includegraphics[keepaspectratio]{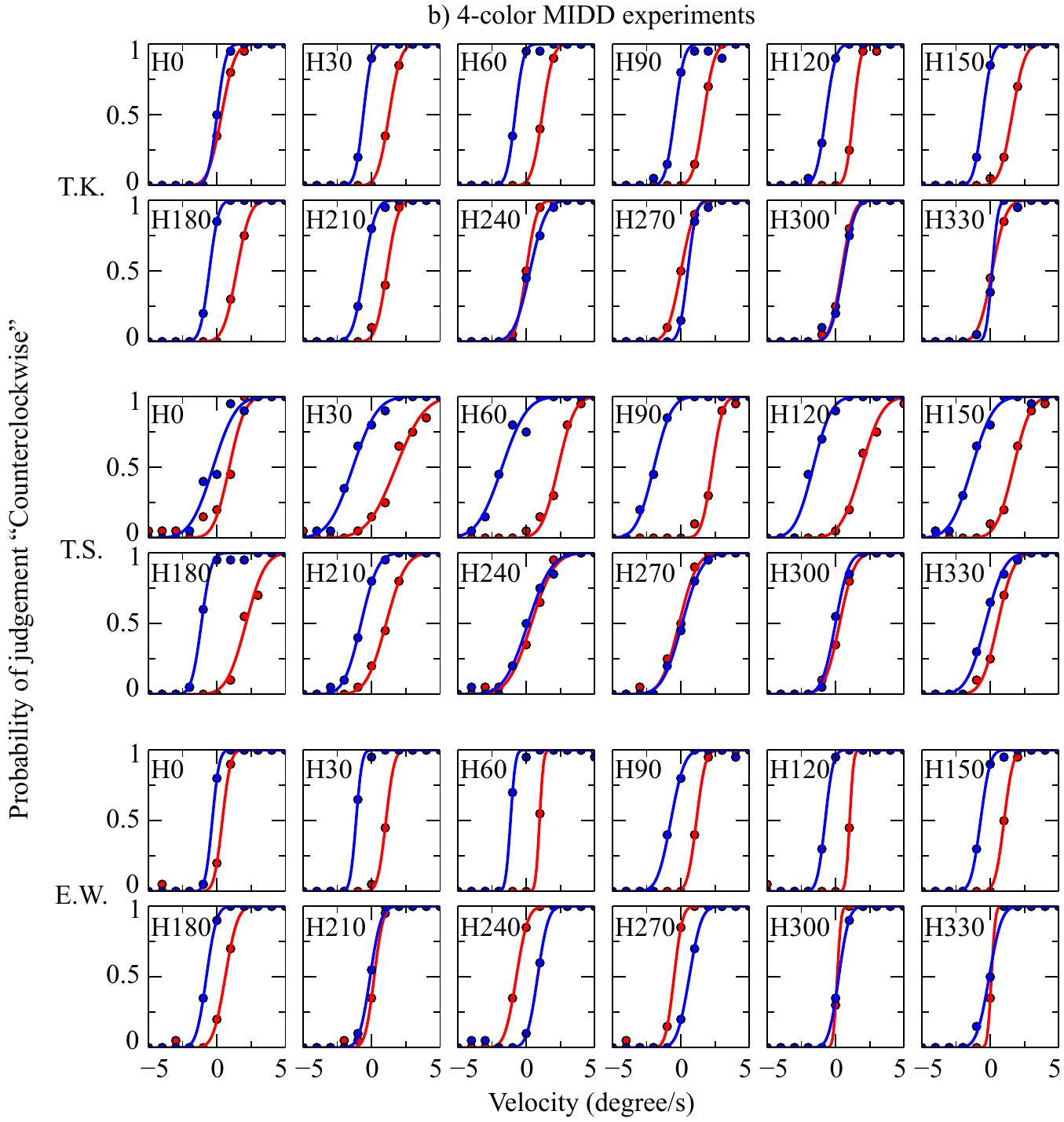}
\caption{Psychometric curves. a) 3-color MIDD and b) 4-color MIDD experiments. The horizontal axis represents the rotational velocity of the stimulus, and the vertical axis shows the probability that the subject responded “it was rotating in the counter-clockwise direction” when the stimulus was presented at that rotation speed. Red and blue are stimuli reversed from each other. The three subjects’ graphs are shown for each varied hue.}
\end{figure}

\end{document}